\documentstyle[12pt,titlepage]{article}

\newcommand{\be}{\begin{equation}}
\newcommand{\ee}{\end{equation}}
\newcommand{\br}{\begin{eqnarray}}
\newcommand{\er}{\end{eqnarray}}

\newcommand{\half}{\frac{1}{2}}
\def\a{{\alpha}}
\def\b{{\beta}}
\def\g{{\gamma }}

\def\m{{\mu}}

\def\o{{\omega}}

\def\dirac#1{\setbox0=\hbox{$#1$}\rlap{\hbox to \wd0{$\hss\mkern1mu/\hss$}} 
\box0 }

\begin{document}

 \begin{titlepage}
\begin{center}

{\bf Q-DEFORMED BOSON OSCILLATORS AND ZERO POINT ENERGY}\\
\vspace{.2in}

\vspace{.2in}

P. Narayana Swamy ($\dagger$ )\\
Department of Physics, Southern Illinois University, Edwardsville, IL 
62026-1654\\
\end{center}
\vspace{.1in}
\begin{center}
{\bf Abstract}
\end{center}

Just as for the ordinary quantum harmonic oscillators, we expect the 
zero-point energy to play a crucial role in the correct high temperature 
behavior. We accordingly  reformulate 
the theory of the statistical distribution function for the q-deformed boson 
oscillators and  develop an approximate theory incorporating the zero-point 
energy. We are then able to demonstrate that for small deformations, the theory reproduces the correct limits both for very high temperatures and 
for very low temperatures. The deformed theory thus reduces to the undeformed 
theory in these extreme cases.

\vspace{.9in}

($\dagger$ ): electronic address: pswamy@siue.edu\\

\vfil
 \end{titlepage}

Let us begin with the classical distribution function of ordinary 
quantum boson harmonic oscillators, when the zero point energy is discarded. 
The occupational probability 
\setcounter{equation}{0}
 \be
P_n = \frac{e^{-\b n \hbar \omega}}{\cal Z}   =   \frac{e^{-\b n \hbar \omega  
}}{\sum_n e^{-\b  n \hbar \omega }} 
\ee
and ${\cal Z}$ the canonical partition function are given by the well-known 
results
\be
 {\cal Z}=\frac{e^x}{ (e^x-1)}, \quad \quad P_n= e^{-nx} (1-e^{-x}),
\ee
where $x= \b \hbar \omega$. The distribution function is
\be
f = \sum_0 ^{\infty} n P_{n}=\left (1- e^{-x}  \right) \sum_0^{\infty} n 
e^{-nx}= \frac{1}{e^x-1} = \frac{1}{e^{\b \hbar \omega} -1},
\label{eq3}
\ee
the familiar result for the ordinary harmonic oscillators. The q-deformed 
theory is described by the $q-$analogue of the algebra of creation 
and annihilation operators of the boson harmonic oscillators, [\ref{Bied}
-- \ref{Nesk} ] defined by 
\be
[a,a] = [a^{\dagger}, a^{\dagger}] =0, \;\; aa^{\dagger}-q a^{\dagger} a= q^{-N}
\ee
where  $q$  is the deformation parameter and $N$ is the number operator which 
obeys the commutation relations
\be
[N,a^{\dagger}]=a^{\dagger},\;\; [N,a]=-a.
\ee
The distribution function  is  [\ref{Songetal}] the $q-$generalization of 
Eq.(\ref{eq3}), thus
\be
f_q = \sum\;  [n] \; P_n
\label{eq1}
\ee
where $[n]$ stands for $n_q$ and is the set of basis numbers, defined by
\be
[\;n\;]= \frac{q^n - q^{-n}}{q - q^{-1}}.
\ee
This is in contrast to the case of the 
ordinary boson oscillators wherein the summation involves $n$, the eigenvalue of 
the number operator. Accordingly many results 
containing the number $n$, eigenvalue of the number operator, have their 
$q-$analog with $n$ replaced by $[n]$. For instance, the Hamiltonian of 
the $q-$boson oscillator is given by
\be
H= \sum_k \left (  \frac{1}{2m}p_k^2 + \half m \omega_k^2 q_k^2  \right 
)=\sum_k \half \hbar \omega_k a^{\dagger}_k a_k   
\ee
and takes the form
\be
H= \sum_k \,  [N] \, \hbar \omega_k  
\label{9}
\ee
which has the eigenvalue
\be
E=\sum_k [n]  \hbar \omega_k  \neq 
\sum_k n \hbar \omega_k.
\ee
With the inclusion of the zero-point energy, the Hamiltonian, Eq.(\ref{9}) 
will be replaced by the following expression 
\be
H= \sum_k \frac{1}{4}  \hbar \omega_k  \left ( 2 a_k a^{\dagger}_k  + 2 
a^{\dagger}_k  a_k \right ) = \sum_k \half \hbar \omega_k \left (  [N+1] + [N] 
\right ).
\label{11}
\ee
We note that $[N]$ is not the number operator. As a consequence, obtaining the distribution function in a closed form for the case of the
$q-$deformed oscillators is no longer as straightforward as in the case of 
ordinary boson oscillators. Herein lies the mathematical difficulty, and 
because of the occurrence of  $[n]$, obtaining a closed form for the 
distribution function does not seem to be feasible.

To deal with this problem there are two methods known in the literature. 
One consists of developing a perturbation theory based on the smallness 
of the deformation parameter $q-1$, or rather in powers of $\g=\ln q$. In 
other words one can calculate the Hamiltonian, the partition function and 
other quantities of interest valid to some order in $\g$. This is the method 
followed by Neskovic and Urosevic [\ref{Nesk}] who derive such perturbation 
theory results for various thermodynamic quantities,  demonstrate how to 
apply this method to some $q-$deformed boson systems, and study typical 
consequences of small deformations.  

The second method consists of an iteration scheme developed by Song et al 
[\ref{Songetal}] which we shall use in this work. This method may be briefly 
reviewed as follows. As a first iteration  one may 
approximate the Partition function occurring in Eq.(\ref{eq1}) by the zeroth 
order result, namely the one corresponding to the Partition 
function of undeformed oscillators, or equivalently the zeroth order 
expression for the occupational probability $P_n$.  We shall henceforword omit the 
subscript $k$. Thus we obtain the first order iteration result for the distribution 
function of the $q-$deformed harmonic oscillator:
\be
f_q = \sum \;[n]\; P_n = \sum_0^{\infty} \frac{q^n - q^{-n}}{q - q^{-1}} 
P_n^{(0)}= 
\frac{1-e^{-x}}{q-q^{-1}}\sum_0^{\infty} \left (  q^n e^{-nx}- q^{-n}e^{-nx}  \right )
\ee
where $x= \b \hbar \omega$. Evaluating the sum and simplifying, we obtain 
\be
f_q= \frac{e^x-1}{(q^{-1}e^x-1) (qe^x-1)} \; =\; \frac{e^{\b \hbar \omega}}{ 
(qe^{\b \hbar \omega} -1)  (q^{-1}e^{\b \hbar \omega} -1)}.
\ee
This is the form derived by Song et al [\ref{Songetal}]. The Grand Partition 
Function which gives rise to this distribution function has been studied 
extensively [\ref{PNS1}, \ref{PNS2}] to derive the thermodynamic properties of 
$q-$bosons as well as in a study of $q-$phonons. This approximate 
distribution has the correct undeformed limit: 
\be
\lim_{q \rightarrow 1}f_q = \frac{1}{e^{\b \hbar \omega} -1}
\ee
It is thus
correct in this sense but unfortunately this distribution function does not exhibit the correct high 
temperature limit. We can suspect that this is due to the fact that the above 
theory has discarded the zero point energy.

Let us first review the well-known high temperature limit of the standard boson 
oscillators in the theory that includes the zero-point energy, where the canonical 
Partition function is given by
\be
{\cal Z}= \sum_{0}^{\infty} e^{-(n+\half)x }= \frac{e^{-\half x}}{1-e^{-x}}= \frac{1}{2 \sinh \half x}
\ee
and the occupational probability is  given by
\be
P_n = \frac{e^{-(n+\half)x}}{\sum_0^{\infty} e^{-(n+\half)x } } \; = 2  
e^{-(n+\half)}x\; \sinh \frac{x}{2}  .
\label{1.15}
\ee
Computing the distribution function, we then find
\be
f= \sum_0^{\infty} (n+\half)P_n \; = 2 e^{-x/2} \sinh \frac{x}{2} \left (   
\half \sum e^{-nx} + \sum n e^{-nx} 
\right )
\ee
and hence
\be
f = \half + \frac{1}{e^{x}-1}\; = \half \;
\left (\frac{e^x +1}{e^x -1} \right ).
\label{1.16}
\ee
For $N$ non-interacting boson oscillators, the internal energy is $N$ times the mean energy of 
one oscillator, thus
\be
U = N \hbar \omega f \; = N \hbar \omega \left (  \half + \frac{1}{e^{\b \hbar 
\omega} -1}  \right ).
\label{1.17}
\ee
In the high temperature limit, $\b \hbar \omega << 1$, we then obtain
\br
( U ) _{kT >> \hbar \omega } \; &=& N \half \hbar \omega + N \hbar \omega 
\frac{1}{1 + \b \hbar \o + \half (\b \hbar \o)^2 + \cdots  -1} \nonumber \\
&\approx & N \half \hbar \omega + \frac{N}{\b}(1- \half \b \hbar \omega + 
\cdots )   = NkT,
\er
the well-known result: the internal energy of quantum 
boson oscillators is the same as the classical value in the high temperature 
limit. This happens [\ref{stocker}] precisely because the term $- \half N \hbar \o $ cancels 
the term arising from the zero-point energy.

It is evidently necessary to include the zero point energy in the theory of 
$q-$deformed boson oscillators and we shall proceed accordingly as follows. We 
see that the distribution function given by  Eq.(\ref{eq1}) gets modified as
\be
f_q = \sum_0^{\infty} \half \left (  [n+1] + [n]  \right )P_n,
\ee
when 
we incorporate the zero-point energy analogous to the Hamiltonian
\be
H = \half \hbar \o \left (  [N+1] + [N] \right ).
\ee
of Eq.(\ref{11}). These have the correct $q \rightarrow 1$ limits. The occupation 
probability  given by the lowest iteration, Eq.(\ref{1.15}), corresponding to the undeformed 
oscillator is thus
\be
P_n^{(0)} = 2 \sinh {\frac{x}{2} } e^{-(n+\half)x}.
\ee
The basis number can be expressed in terms of $q=e^{\g}$:
\be
[n] = \frac{q^n - q^{-n}}{q - q^{-1}} \; = \; \frac{\sinh n \g}{\sinh \g}.
\ee
We can thus express the distribution function as
\be
f_q = \sum_0^{\infty} \frac{\sinh n \g + \sinh (n+1)\g}{\sinh \g} \sinh \frac{ 
x}{2} 
e^{-(n+\half) x}.
\ee
Evaluating the sum and simplifying the algebra, we obtain 
\be
f_q =  \frac{1-e^{-x}}{2(q-q^{-1})} \left (  \frac{e^x}{e^x-q} - 
\frac{e^x}{e^x-q^{-1}} + \frac{q e^x}{e^x-q} -  \frac{ q^{-1}e^x}{e^x-q^{-1}}
  \right )
\ee
which can be conveniently put in the form
\be
f_q = \half \frac{(e^{2 \b \hbar \o} -1)}{ (e^{\b \hbar 
\o}-q) (e^{\b  \hbar \o} -q^{-1})},
\ee
or alternatively,
\be
f_q = \half  \frac{(e^{2 \b \hbar \o} -1)}  
{(q e^{\b \hbar \o} -1) (q^{-1}e^{\b \hbar \o}-1 ) 
}.
\ee
It is easily verified that in the limit $q \rightarrow 1$, the above goes over to 
the form for the standard boson oscillator with the inclusion of the zero 
point energy, given by Eq.(\ref{1.16}). It is convenient to rewrite the 
distribution function in terms of partial fractions, thus 
\be
f_q = \half \left (  \frac{C_1}{1- q e^{-x}} + \frac{C_2}{1- q^{-1} e^{-x}}+ 
\frac{C_2 }{q^{-1} e^{x} -1}+ \frac{C_1 }{q e^{x}-1}
  \right ),
\ee
where $C_1 + C_2 =1$, and
\be
C_1 = \frac{q}{q-q^{-1}}, \;\; C_2= - \frac{q^{-1}}{q-q^{-1}}.
\ee
It is expedient at this point to introduce the  chemical potential $\m$ 
primarily to 
facilitate some manipulations, and we shall set $\m=0$ later at the end. Thus we make the replacement
\be
e^x= e^{\b (E-\m)}= \frac{1}{z} e^{\b E}, \quad e^{-x}= e^{-\b (E-\m)}= 
{z}e^{\b E},
\ee
where $z=e^{\b \m}$ is the fugacity. Hence we obtain
\be
f= \half \left (  \frac{C_1}{1- q z e^{-\b E}} + \frac{C_2}{1- \frac{z}{q}   
e^{-\b E}}
+ \frac{C_2}{1- \frac{1}{qz}   e^{-\b E}}  + \frac{C_1}{\frac{q}{z}  
 e^{\b E}-1} \right ).
\label{1.32}
\ee

We may now introduce the first iteration Partition Function ${\cal Z}$ which corresponds to 
this distribution function. We accordingly determine
\be
{\cal Z}\; = \; \prod_E \left (  1-\frac{1}{qz}e^{\b E}  \right )^{-\half C_1}  
\; \left (  
1-\frac{q}{z}e^{\b E}  \right )^{-\half C_2} \; \left (  {qz}e^{-\b E}-1  
\right )^{-\half C_2} \; \left (  \frac{z}{q}e^{-\b E}-1  \right )^{-\half C_1}
\ee
as the first iteration Partition Function. We may compute the thermodynamic 
potential for the $q-$boson oscillator as
\br
\Omega & = & - \frac{1}{\b} \ln {\cal Z}\nonumber\\
&=&  \sum_E\frac{1}{2\b} \left \{ C_1 \ln \left (  1- \frac{1}{qz}e^{\b E} 
\right )
 +C_2 \ln \left (  1- \frac{q}{z}e^{\b E}  \right ) \right. \nonumber \\
\hspace*{.8in} & +& \left.  C_2 \ln \left (  {qz}e^{-\b E}-1  \right 
) +
 C_1 \ln \left (  \frac{z}{q}e^{-\b E}-1  \right )
 \right \}. 
\er
If we now compute the distribution function
\be
f_q = - \b z \frac{\partial}{\partial z} \Omega ,
\ee
we readily reproduce 
the distribution in Eq. ({\ref{1.32}}). At this point we 
may, for simplicity, return to the case where the chemical potential is 
zero. Thus dropping $\m$, we obtain the distribution function
\be
f_q = \half \sum \left (  \frac{C_1}{ 1- q e^{-\b E}} + \frac{C_2} {1- q^{-1} 
e^{-\b E}}   + \frac{C_2}{ q^{-1} e^{\b E}-1  }  +\frac{C_1}{  q e^{\b 
E}-1}\right )
\ee
and the Partition Function
\be
{\cal Z}= \prod_E \left (  1-q^{-1}e^{\b E}  \right )^{-\half C_1} \left (  
1-q e^{\b E}  \right )^{-\half C_2} \left ( q e^{-\b E}-1  \right )^{-\half 
C_2}  \left (  q^{-1}e^{-\b E}-1  \right )^{-\half C_1}.
\ee
It is readily verified that these forms possess the correct $q \rightarrow 1$ 
limits, Eq.(\ref{1.16}) and 
\be
{\cal Z}=\prod \frac{1}{2 \sinh {\frac{x}{2}}  }.
\ee
 We may now compute the internal energy: 
\br
U&=& - \frac{\partial}{\partial \b}\ln {\cal Z}\\
&=& -\half \sum E \left (  \frac{C_1}{ q e^{-\b E}-1} + \frac{C_2} { q^{-1} 
e^{-\b E}-1}   + \frac{C_2}{1- q^{-1} e^{\b E}  }  +\frac{C_1}{1-  q e^{\b 
E}}\right ),
\er
which clearly has the correct $q \rightarrow 1$ limit, 
namely
\be
U_{q=1}= \sum E \left (  \half + \frac{1}{e^x-1}  \right ).
\ee

It is convenient to express the internal energy in an alternate form, since the internal 
energy is $N$ times the mean energy of one oscillator,
\be
U= \half N\,E  \left (  \frac{C_1}{ 1- q e^{-\b E}} + \frac{C_2} {1- q^{-1} 
e^{-\b E}}   + \frac{C_2}{ q^{-1} e^{\b E}-1  }  +\frac{C_1}{  q e^{\b 
E}-1}\right ),
\ee
The undeformed limit is
\be
U_{q=1}= - \half N E \frac{\sinh \b E}{1 - \cosh \b E} = \half N E \frac{e^x + 
1}{e^x - 1}
\ee
as expected. For $q\not=1$, we may 
cast it in the form
\be
U= -NE \frac{\sinh \b E}{q-q^{-1}} \left (  \frac{q^2}{1 + q^2 - 2q \cosh \b 
E} - \frac{q^{-2}}{1 + q^{-2}-2q^{-1} \cosh \b E  }
  \right ).
\ee
Upon introducing $q=e^{\g}$, 
we derive the result in a convenient form
\be
U= NE \frac{\sinh \b E } {2 (\cosh \b E - \cosh \g) } \; = \half N E \frac{ 
\sinh { \frac{ \b E}{2}} \;  \cosh { \frac {\b E} {2} } } {\sinh  {\frac {\b E 
+ \g} {2}} 
\; \sinh {\frac {\b E - \g}{2}} }
\label{44}
\ee
which has the familiar $q \rightarrow 1$ limit, namely
\be
U_{q=1}= \half N E \frac{1}{\tanh {\frac{\b E} {2} }    }.
\label{45}
\ee 

Let us first consider the low temperature limit. For $\b E >>1$, the internal 
energy will not have the correct low temperature limit for arbitrary $q$ since 
it will depend on $\g$. We shall thus consider small deformations only, $\g << 
1$, or rather $\g << \b E$ in terms of the dimensionless quantity.  
For very low temperatures, we thus have $\b E >> 1 >> \g$. In this case then
\be
U_{\b \rightarrow \infty}= \lim_{\b \rightarrow \infty} \half N E \frac{\sinh \b E/2 \cosh \b E/2}{(\sinh 
\b E/2)^2} = \half N E 
\ee
which is the zero-point-energy. The low temperature limit 
is the pure quantum effect as expected.  Thus the theory at $T = 0$ is the 
same as the undeformed theory [\ref{Birman}] .

Next we examine the high temperature limit. In the case $q=1$ , it has the correct classical limit, 
namely
\be
U_{q=1} \rightarrow N k T, \;\; \b E << 1
\ee
but this is not the case when $q \not=1$ for arbitrarily large values of $\g 
\not=0$. We accordingly consider small deformations defined by $\g << \b E$. At 
high temperatures we thus have $\g << \b E << 1$. Consequently we find from 
Eq.(\ref{44}) that

\be
U_{\b \rightarrow 0} = \half NE \frac{2}{\b E}= N \b,
\ee
which is the expected classical limit. This implies that as long as the 
deformations are small, the high temperature limit of the $q-$deformed boson 
oscillators is the same as that of the undeformed theory. It is gratifying to 
note that this is parallel to the expectation that at $T = 0$, the theory 
of the $q-$deformed oscillators is the same as that of the undeformed theory 
[\ref{Birman}].

In summary, we have developed the theory of the statistical mechanics of 
$q-$deformed boson oscillators incorporating the zero-point-energy. We have 
confronted the theory against the correct limits at very high temperatures and 
very low temperatures. we thus 
conclude that (a) At very low temperatures, $T =0$, the theory is no different 
from the undeformed theory of boson oscillators and (b) at very high 
temperatures, $T = \infty$, the internal energy corresponds to the classical 
limit and the theory reduces to the undeformed case at $T = \infty$ and (c) 
these correct limits are possible only after incorporating the zero-point 
energy in the spectrum.  

The theory of $q-$deformed fermion oscillators pose different challenges. This 
investigation will be reported in a future publication.

{\bf Acknowledgments}
\vspace{.2in}

I am grateful to P. Quarati and A. Lavagno of Politecnico di Torino, Torino,  
Italy for the kind hospitality where this work was begun. I am thankful to A. 
Lavagno for valuable discussions and drawing my attention to some papers on 
the subject of $q-$oscillators. I thank R. 
Ramachandran, director of the Institute of Mathematical Sciences (IMSc), 
Chennai, India, for the hospitality extended to me, where I did much of this 
work during part of my sabbatical leave in 1997. I would like to thank colleagues at 
IMSc: G. Rajasekaran for many valuable discussions and for educating me on
the nuances of the $q-$deformation algebra; R. Parthasarathy for discussions 
on the general subject of $q-$oscillators, and R. Jagannathan and G. Baskaran for 
stimulating discussions. 

\vspace{.2in}

{\bf REFERENCES AND FOOTNOTES}
\vspace{.2in}

\begin{enumerate}

\item \label{Bied}L. Biedenharn, {\it J. Phys.\/} {\bf A22}, L873 (1989)

\item \label{lit} Q.Yang and B.Xu, {\it J. Phys. \/}{\bf A 26}, L365 (1993); 
N. Aizawa, {\it ibid\/} {\bf A26}, 1115 (1993);  K. Viswanathan {\it et al 
ibid\/}, {\bf A25}, L335 (1993); E. G. Floratos, {\it J. Phys.\/} {\bf A24}, 
4739-4750 (1991); T. Altherr and T. Grandou, {\it Nuc. Phys.\/} {\bf B402}, 
195-216 (1993)

\item \label{Partha} R. Parthasarathy and K. S. Viswanathan, {\it J. Phys.\/} 
{\bf A 24}, 613-617 (1991), K. S. Viswanathan, R. Parthasarathy and R. 
Jagannathan, {\it ibid\/}{\bf A25}, L335-339 (1992).

\item \label{Graj} A.K. Mishra and G. Rajasekaran, {\it Mod. Phys. Lett.\/} 
{\bf A 9}, 419-426 (1994); A. K. Mishra and G. Rajasekaran, {\it Pramana\/} {\bf 45}, 
91--139 (1995). 

\item \label{Songetal} H.Song, S.Ding and I.An, {\it J. Phys. \/} {\bf A  26}, 
5197 (1993).

\item \label{Lavagno} G. Kaniadakis and P. Quarati, {\it Phys. Rev.\/} {\bf 
E49}, 5103 (1994); G. Kaniadakis, {\it Phys.Rev.\/} {\bf E49}, 5111 (1994); G. 
Kaniadakis, A. Lavagno and P. Quarati, Nucl. Phys. {\bf B 466} 527 (1996). 
{\it ibid\/}, {\it Phys. Lett. \/}{\bf A 227} 227 (1997). 

\item \label{Nesk}P. Neskovic and B. Urosevic, {\it Int. Journal Mod. 
Phys.A\/}{\bf A7}, 3379 (1992)

\item \label{PNS1} P. Narayana Swamy, {\it Int. Journ. Mod. Phys.\/} {\bf B 
10}, 683-699 (1996).

\item \label{PNS2} P. Narayana Swamy, {\it Mod. Phys. Lett.\/} {\bf B 10}, 
23-28 (1996).

\item \label{stocker} W. Greiner, L. Neise and H. Stocker, {\it Thermodynamics 
and Statistical Mechanics\/}, Springer-Verlag  Inc. (New York) 1995.

\item \label{Birman} J. L. Birman, Phys. Lett. {\bf A 167}, 363 (1992). That 
$q-$deformation does not prevail at $T=0$ is also apparent in a 
$q-$deformed theory of fermionic oscillators. See {\it e.g.,\/} C. R. Lee 
and J. P. Yu, Phys. Lett. {\bf A 164}, 164 (1992). 

\end{enumerate}

\end{document}